\documentclass[sigconf]{acmart}
\AtBeginDocument{%
  \providecommand\BibTeX{{%
    \normalfont B\kern-0.5em{\scshape i\kern-0.25em b}\kern-0.8em\TeX}}}

\copyrightyear{2024}
\acmYear{2024}
\setcopyright{rightsretained}
\acmConference[Preprint]{Preprint}{July 2024}{Toronto, Canada}
\begin{document}

\title{Understanding Help-Seeking Behavior of Students Using LLMs vs. Web Search for Writing SQL Queries}


% Harsh Kumar, Mohi Reza, Jeb Mitchell, Ilya Musabirov, Lisa Zhang and Michael Liut

\author{Harsh Kumar\textsuperscript{*}} % Harsh
\orcid{0000-0003-2878-3986}
\affiliation{
\institution{University of Toronto}
   \city{Toronto}
   \country{Canada}
}
\email{harsh@cs.toronto.edu}

\author{Mohi Reza\textsuperscript{*}} % Mohi
\orcid{0000-0001-9668-3384}
\affiliation{
\institution{University of Toronto}
   \city{Toronto}
   \country{Canada}
}
\email{mohireza@cs.toronto.edu}

\author{Jeb Mitchell} % Jeb
\affiliation{
\institution{University of Toronto}
   \city{Toronto}
   \country{Canada}
}
\email{email@mail.utoronto.ca}

\author{Ilya Musabirov} % Ilya
\orcid{0000-0003-2965-5302}
\affiliation{
\institution{University of Toronto}
   \city{Toronto}
   \country{Canada}
}
\email{imusabirov@cs.toronto.edu}

\author{Lisa Zhang} % Lisa
\orcid{0000-0002-7302-6530}
\affiliation{
\institution{University of Toronto Mississauga}
   \city{Mississauga}
   \country{Canada}
}
\email{lczhang@cs.toronto.edu}

\author{Michael Liut} % Michael
\orcid{0000-0003-2965-5302}
\affiliation{
\institution{University of Toronto Mississauga}
   \city{Mississauga}
   \country{Canada}
}
\email{michael.liut@utoronto.ca}

\thanks{\textsuperscript{*}Both authors contributed equally to this research.}

\renewcommand{\shortauthors}{Kumar and Reza, et al.}
% \input{authors_anon}

\begin{abstract}
Growth in the use of large language models (LLMs) in programming education is altering how students write SQL queries. Traditionally, students relied heavily on web search for coding assistance, but this has shifted with the adoption of LLMs like ChatGPT. However, the comparative process and outcomes of using web search versus LLMs for coding help remain underexplored. To address this, we conducted a randomized interview study in a database classroom to compare web search and LLMs, including a publicly available LLM (ChatGPT) and an instructor-tuned LLM, for writing SQL queries. Our findings indicate that using an instructor-tuned LLM required significantly more interactions than both ChatGPT and web search, but resulted in a similar number of edits to the final SQL query. No significant differences were found in the quality of the final SQL queries between conditions, although the LLM conditions directionally showed higher query quality. Furthermore, students using instructor-tuned LLM reported a lower mental demand. These results have implications for learning and productivity in programming education.
\end{abstract}

\begin{CCSXML}
<ccs2012>
   <concept>
       <concept_id>10010405.10010489.10010490</concept_id>
       <concept_desc>Applied computing~Computer-assisted instruction</concept_desc>
       <concept_significance>500</concept_significance>
       </concept>
   <concept>
       <concept_id>10003120.10003121.10011748</concept_id>
       <concept_desc>Human-centered computing~Empirical studies in HCI</concept_desc>
       <concept_significance>500</concept_significance>
       </concept>
   <concept>
       <concept_id>10003456.10003457.10003527.10003539</concept_id>
       <concept_desc>Social and professional topics~Computing literacy</concept_desc>
       <concept_significance>500</concept_significance>
       </concept>
 </ccs2012>
\end{CCSXML}

\ccsdesc[500]{Applied computing~Computer-assisted instruction}
\ccsdesc[500]{Human-centered computing~Empirical studies in HCI}
\ccsdesc[500]{Social and professional topics~Computing literacy}

\keywords{Large Language Models, Human-Computer Interaction, SQL Query-writing, Experiments, Data Systems Education}

\begin{teaserfigure}
  \includegraphics[width=\textwidth]{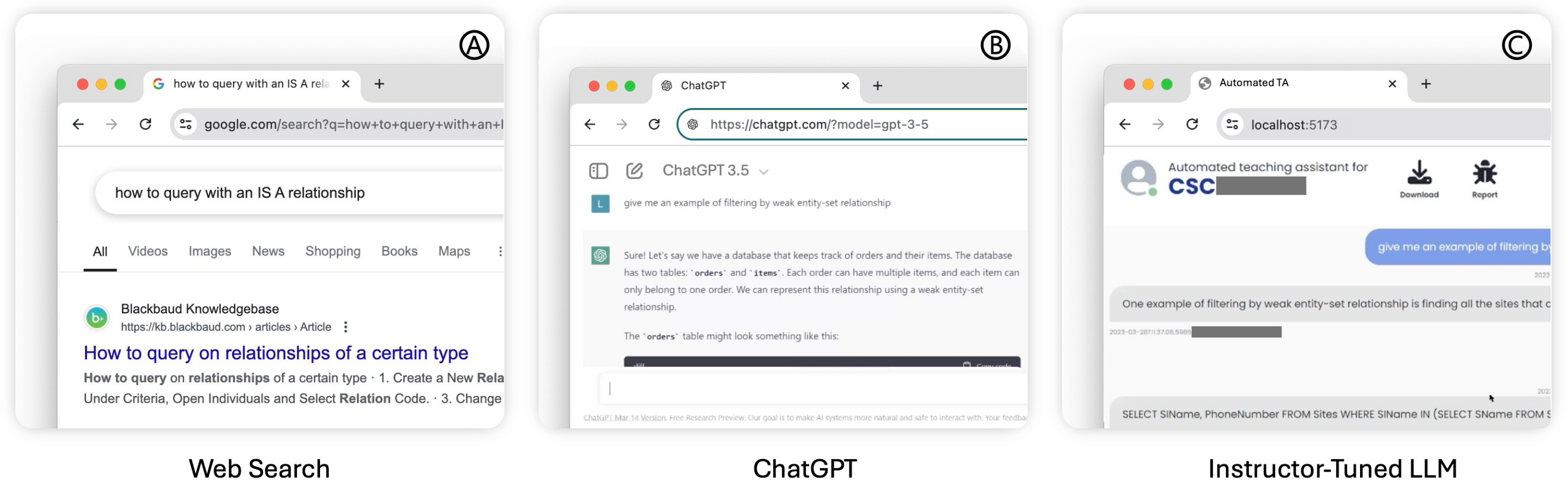}
  \caption{Through a randomized interview study in an SQL course, we compared how students use web search (A) with an off-the-shelf LLM (B) and an instructor-tuned LLM chatbot (C) that has knowledge about the course context and content.}
  \Description{Figure comparing the three study conditions: web search, chatgpt, and an instructor-tuned LLM.}
  \label{fig:teaser}
\end{teaserfigure}


\maketitle

\section{Introduction}
%State of the world...
A recent survey \cite{Daines-Hutt_2023} of over three thousand programmers revealed that 84\% are using AI tools, with ChatGPT being the most popular—74.9\% of developers use it on a weekly basis. The most popular use-case (over 80\%) is using these tools as a \textit{search engine} for new topics, highlighting the potential for LLM-based chatbots to supplement the programming learning process alongside conventional search methods. For structured tasks such as formulating SQL queries, LLM chatbots offer the unique ability to generate tailored responses coupled with explanations. This contrasts with traditional search methods, where learners must hunt for and adapt code snippets from sites like Stack Overflow.

%The big BUT...
Despite the growing popularity of LLM chatbots for search-engine-like tasks, the comparative impact of traditional web search versus LLM chatbots in real-world programming classrooms, particularly in supplementing student engagement and learning of languages like SQL, is not well-understood. Empirical insights into how students engage with these tools can inform the design of better learning support tools, especially in flipped classroom settings where self-regulated learning through out-of-class practice plays an important role. Furthermore, exploring whether tuning off-the shelf chatbots like ChatGPT using cost-effective and easy methods, such as adding system prompts, can enhance their usefulness for students is an open question that is of interest to many instructors who are navigating the integration of AI tools into their teaching, while trying to mitigate many of the common concerns that off-the-shelf LLM chatbots pose, like their tendency to generate direct answers, or their lack of awareness of course specific content and context. 

%Therefore we did...
To address these issues, we conducted a mixed-methods randomized interview study with 39 students where we compared traditional search (e.g., Google, Bing) with both standard ChatGPT (3.5 model, which was the frontier at the time of the experiment) and an instructor-tuned version of the chatbot with added guard rails (e.g. being told not to give out direct answers) and course context (e.g. a description of the learning goals and content covered by the course). We asked two research questions:

\begin{itemize}
    \item \textbf{RQ1:} How does traditional web search (e.g., Google, Bing) compare to LLM-based chatbots like ChatGPT in terms of student engagement and learning outcomes in programming education, particularly for languages like SQL?
    \item \textbf{RQ2} Can low-cost tuning methods, such as adding system prompts, be employed by instructors to enhance the effectiveness of LLM-based chatbots like ChatGPT for educational purposes, and if so, how does this tuning influence student engagement and learning outcomes?
\end{itemize}

We found that students interacted with the instructor-tuned LLM more than twice as much compared to both standard ChatGPT ($p = 0.01$) and web search ($p < 0.0001$). Despite this increased engagement, there were no significant differences in the correctness of the final SQL queries across conditions, although the LLM conditions showed higher query quality directionally. These results suggest the potential value of domain experts tuning LLMs using inexpensive methods like system prompts to enhance learner engagement.

The main contributions of this work are:
\begin{itemize}
    \item Findings from a randomized experiment in a real-world SQL classroom comparing web search with plain ChatGPT and an instructor-tuned LLM, demonstrating how instructor-tuning can significantly impact engagement.
    \item A useful snapshot of GPT-3.5 usage for SQL education, serving as a benchmark for future studies with newer models.
    \item Empirical insights into web-search and LLM chatbot usage that can inform the design of learning and productivity tools for programming education using AI.

\end{itemize}

\section{Related Work}

The advent of large language models (LLMs) has prompted comparative research with traditional web search methods for information retrieval and problem solving. Recent studies have compared web search and LLM querying for general information-seeking tasks. Wazzan \textit{et al.} found that web search outperformed LLMs in accuracy in a geolocation task, with LLM users struggling to formulate effective queries \cite{Wazzan24}. Xu observed that LLM users spent less time on their tasks and demonstrated more consistent performance across education levels, but accuracy suffered in fact-checking and complex tasks compared to participants using web search \cite{Xu23}. Spatharioti \textit{et al.} reported that LLM users completed product comparison tasks more quickly by using fewer, more complex queries \cite{Spatharioti23}. While LLM users generally reported higher satisfaction and perceived response quality \cite{Xu23, Spatharioti23}, their accuracy was dependent on the reliability of LLM-provided information and effective prompting \cite{Xu23, Spatharioti23}. 

Despite widespread adoption of LLMs in programming tasks \cite{guo2023six, dongback, kazemitabaar2024codeaid, prather2023robots, denny2024computing}, fewer comparative studies have been conducted between web search and LLM querying in computer science and computer science education. Research shows that professionals select between web search and querying LLMs through search strategies that utilize self-reflection on their knowledge of the problem and domain \cite{Yen24}. Yen \textit{et al.} found that web search is preferred when professionals are unfamiliar with the domain, or the problem is poorly-defined, because it returns greater diversity of results than LLM queries. Using an LLM is preferred when the user believes the problem is discussed frequently enough that the model has awareness of it, and they possess sufficient knowledge to identify incorrect responses \cite{Yen24}. In contrast, research into students' use of web-based resources showed that students often exhibit shallow, trial-and-error approaches without clear strategy or self-reflection \cite{Wong24}. While some students use web search for general reference \cite{Skripchuk24}, other students tend to seek quick answers and exact code matches to specific problems \cite{Wong24}. They tend to exhibit a production bias where they focus on solving the immediate task by quick web searches rather than engaging with foundational concepts in course resources \cite{Wong24}. LLMs provide a potential solution to these strategy and orientation problems. Students view LLMs as providing more specific, easily understandable responses \cite{Skripchuk24} and utilize them for various programming education tasks like generating practice exercises, clarifying error messages, or providing tips on syntax \cite{Cambaz24}.

However, for programming novices, both approaches have potential pitfalls. Web searches can lead to the faulty integration of poorly-understood code, resulting in compounding errors \cite{Skripchuk23}. Students’ limited technical vocabulary impedes their ability to define the problem sufficiently for effective keyword-based searches \cite{Wong24}. Students often rely on complex resources like StackOverflow, which they may find difficult to comprehend as novices \cite{Wong24}. With LLMs, there are concerns about overreliance potentially hindering the development of critical thinking and problem-solving skills \cite{Cambaz24}. Students express beliefs that using LLMs is “closer to cheating” and “doesn’t really teach [them] anything” \cite{Skripchuk24}. Additionally, LLM users may become stuck iterating on prompts that produce incorrect results \cite{Yen24}.

Despite the challenges, LLMs show promise in enhancing students' understanding of programming concepts, if their reliability can be improved \cite{jurenka2024towards}. They offer interactive, beginner-friendly explanations and can provide tailored support for various aspects of programming education \cite{Cambaz24, Skripchuk24}. However, there is a clear need for more research directly comparing student use of web search and LLMs in the context of computer science education. This research will enable a more effective integration of these technologies and inform future pedagogical approaches.

\section{Methods}
We conducted a mixed-methods randomized interview study with 39 students where we compared traditional web search with both standard ChatGPT (3.5 model, which was the frontier at the time of the experiment) and an instructor-tuned version of the chatbot with added guard rails (e.g. being told not to give out direct answers) and course context (e.g. a description of the learning goals and content covered by the course). Students were provided Entity-Relationship diagrams and then asked to solve two SQL-writing problems, one after the other, with one of the randomly selected source of help for each question. 

\subsection{Context of Deployment}
This research study occurred in a 12-week introductory database systems course at a large, publicly funded, research-intensive university in North America, and received ethics board approval (Ethics Protocol \#123456\footnote{redacted for review.}) This course targeted third-year undergraduate students in a four-year honors computing program, with $226$ students enrolled in the course. The course used a flipped classroom approach, which allowed students to engage with video content and exercises on a custom Learning Management System (LMS) before attending synchronous in-person lecture sessions. Given the flipped approach, students often sought external resources such as web search and LLM-based chatbots to supplement their self-regulated learning, making this an ideal context for deploying the study.

The study was advertised as an optional activity to test the effectiveness of an LLM-based chatbot tutor. 39 students volunteered to participate and were then enrolled in the interview study. The interviews were conducted over Zoom and lasted almost 30 minutes (each round was time-boxed to 15 minutes). During the interview study, each student was randomly assigned to interact with two sources of help, one after the other, to solve two different types of SQL writing problems (randomized in order). 21 were assigned to the Instructor-tuned LLM vs. Web Search round, while the remaining (18) participated in the Instructor-tuned LLM vs. ChatGPT round. 

\subsection{Experimental Conditions}
This study considered three distinct sources of help in writing code.

\subsubsection{Web Search} Students in this round could freely use any web search engine to find resources to help write the SQL query. All students used Google search and then navigated to coding blogs and forums such as GeeksForGeeks, StackOverflow, etc. Figure \ref{fig:teaser}A shows an example query issued by one of the students in the study.

\subsubsection{ChatGPT} In this round, students were directly given access to ChatGPT (3.5). No additional tweaks were made to the model. This acted as a proxy for publicly available LLMs being used by students to solve assignment problems on their own. Figure \ref{fig:teaser}B shows a student's interaction with ChatGPT during the study.

\subsubsection{Instructor-tuned LLM} Students in this round were given access to a chatbot using GPT-3.5 (the frontier model at the time of the study). The model was configured with a system prompt to be particularly helpful for writing the SQL queries and provided the context of the questions used in the study. This acted as a proxy for instructor-provided LLM chatbots used in classrooms (e.g., KhanMigo, CS50 bot \cite{liu2024teaching}, etc.) Figure \ref{fig:teaser}C shows the chatbot interface used in the study.

\subsection{Outcome Measures}
Each interview study session was recorded on video, and the following outcome measures were manually extracted from the videos.

\subsubsection{Number of Interactions with the Source of Help}
We measure this by calculating the number of queries sent to the assigned source of help. Each query constitutes an interaction. 

\subsubsection{Number of Edits Made to the Final SQL Query}
This was measured by calculating the number of changes made to the final SQL query submitted by the student, during each round of the study. 

\subsubsection{Quality of the SQL Query}
We utilized a grading rubric specified by the course instructor for the two types of questions. Based on this rubric, one of the authors assigned a score (0-100) to each of the final SQL queries of the students.

\subsubsection{Self-reported Mental Demand}
We used the \textit{Mental Demand} subscale from the NASA-TLX questionnaire \cite{rubio2004evaluation} to compare the mental demands of completing the SQL-writing task with their assigned source of help. On a scale of 1 (very low) to 10 (very high), students were asked to rate how mentally demanding the task was.

\section{Results}

\begin{figure*}[ht]
    \includegraphics[width=\textwidth]{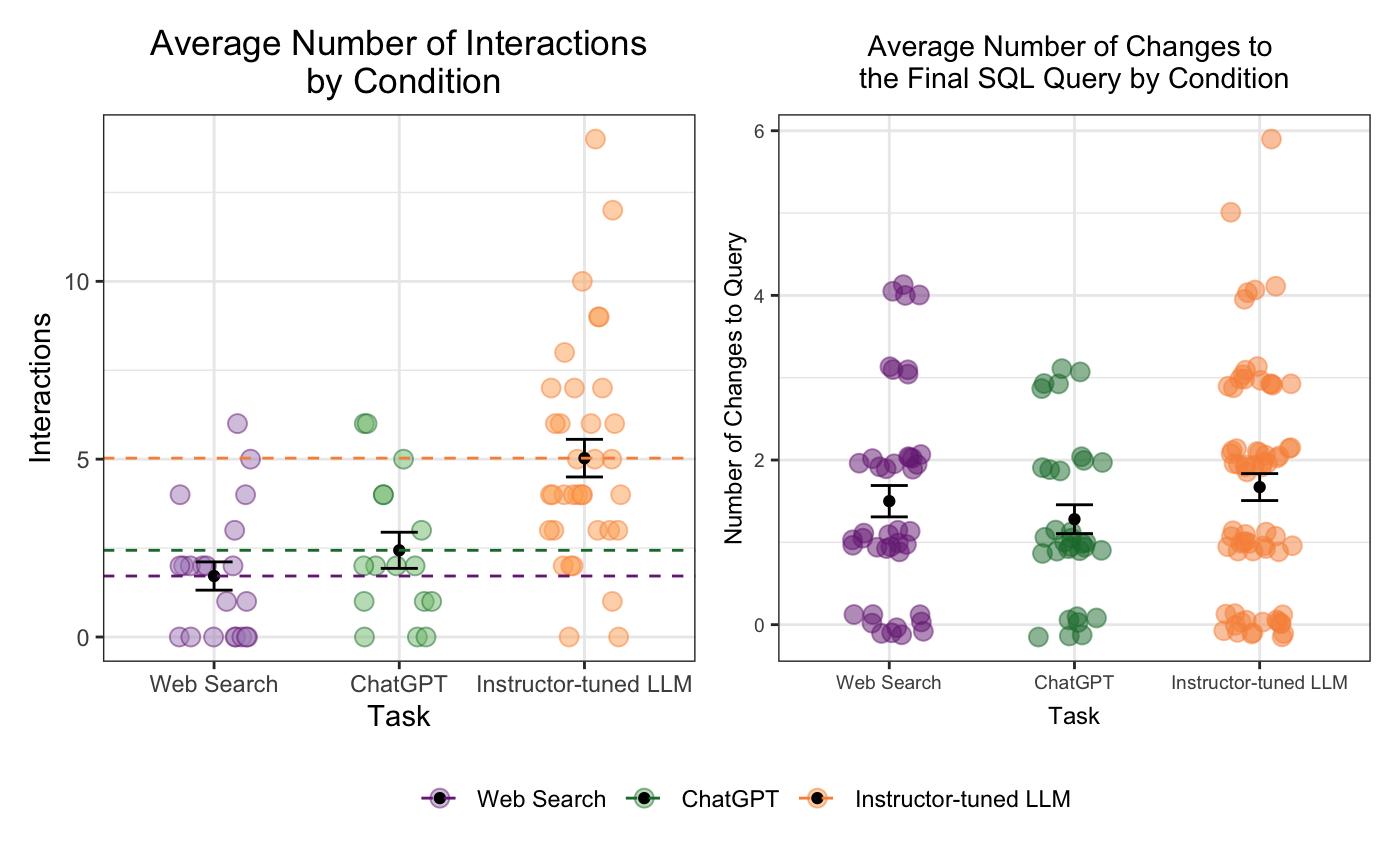}
    \caption{Comparative Analysis of Number of Interactions and Number of Changes to SQL Query between Conditions. The left panel shows the average number of interactions, indicating a higher number of interactions for the Instructor-tuned LLM condition. The right panel evaluates the average number of changes made to the final SQL query, with no significant differences between conditions. Error bars represent +- one standard error of the mean.}
    \label{fig:interaction-change}
\end{figure*}

\subsection{Effect on Number of Interactions}
Figure \ref{fig:interaction-change} left-facet shows the average number of interactions with the assigned source of help by condition. We performed a Kruskal-Wallis H test to compare the number of interactions across the three conditions and the results indicated a statistically significant difference between the groups ($\chi^2(2) = 20.5$, $df = 2$, $p < 0.0001$). Post-hoc pairwise comparisons were conducted using Dunn's test with Bonferroni correction. Students interacted with the Instructor-tuned LLM more than twice compared to ChatGPT ($p = 0.01$) and Web Search ($p < 0.0001$).

% > mean_interactions
% # A tibble: 3 × 3
%   Task                 mean_interactions    se
%   <fct>                            <dbl> <dbl>
% 1 ChatGPT                           2.44 0.508
% 2 Web Search                        1.71 0.397
% 3 Instructor-tuned LLM              5.03 0.529

% \begin{itemize}
%     \item ChatGPT vs. Instructor-tuned LLM: $p = 0.0127$
%     \item ChatGPT vs. Web Search: $p = 1.0000$
%     \item Instructor-tuned LLM vs. Web Search: $p < 0.0001$
% \end{itemize}

\subsection{Effect on Number of SQL Query Edits/Changes}
Figure \ref{fig:interaction-change} right-facet shows the average number of changes made to the final SQL query provided by the student. The Kruskal-Wallis H test for changes between conditions was not significant ($\chi^2(2) = 1.55$, $df = 2$, $p = 0.46$). In all conditions, students needed just under 2 changes to arrive at their final query in the assigned time. However, there were students in the Instructor-tuned LLM condition who made over 5 changes in their final SQL query (Figure \ref{fig:interaction-change}).

% > mean_changes
% # A tibble: 3 × 3
%   Task                 MeanChanges    SE
%   <fct>                      <dbl> <dbl>
% 1 ChatGPT                     1.28 0.175
% 2 Web Search                  1.5  0.191
% 3 Instructor-tuned LLM        1.67 0.164

\begin{figure*}[ht]
    \includegraphics[width=\textwidth]{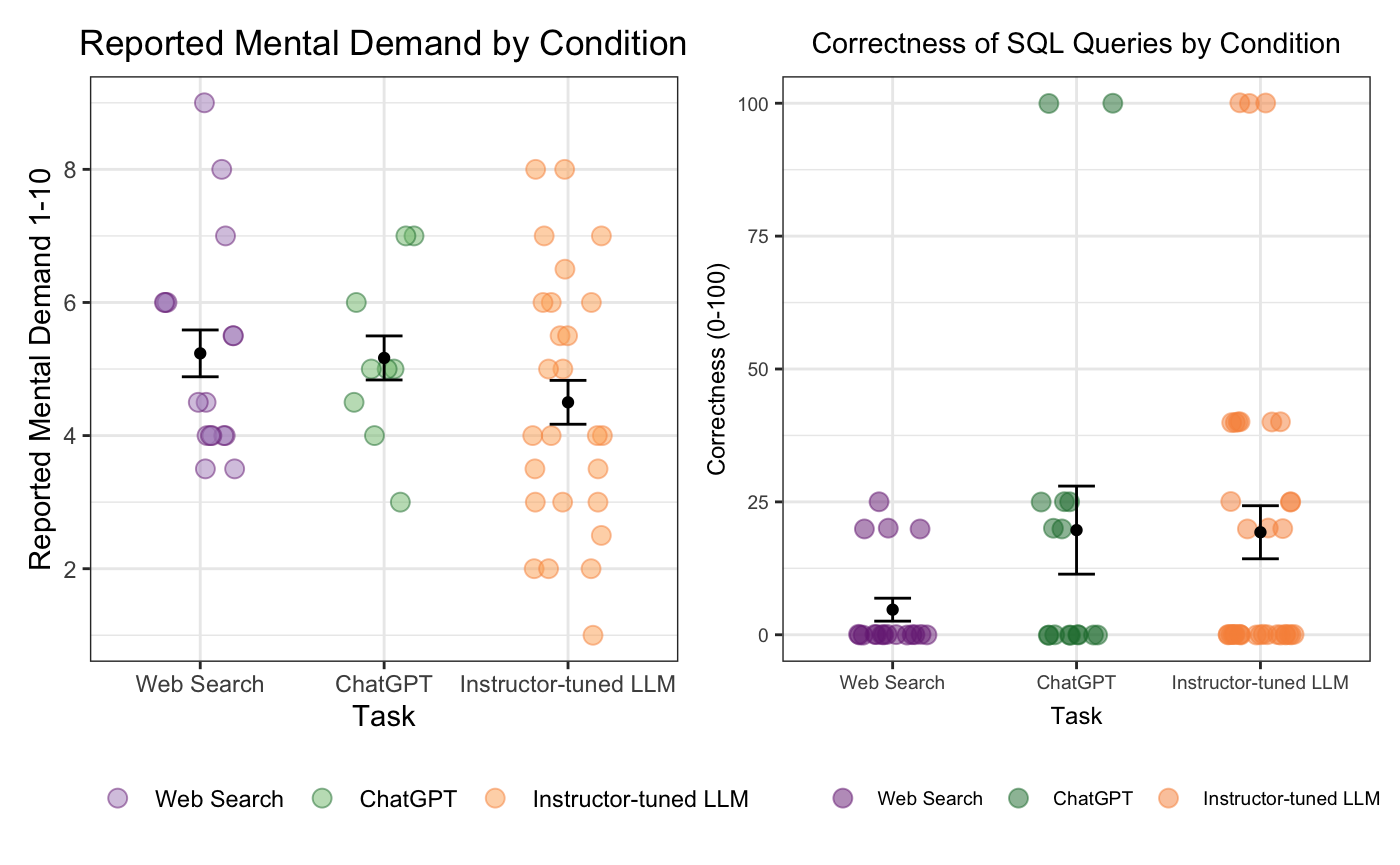}
    \caption{Comparative Analysis of Students' Self-Reported Mental Demand during Task and Correctness of SQL Queries between Conditions. The left panel shows the average reported mental demand, with no significant differences between conditions but directionally lesser average mental demand for students using Instructor-tuned LLM. The right panel shows the average correctness of SQL queries, highlighting higher correctness when students used either of the LLMs compared to the web search condition. Error bars represent +- one standard error of the mean.}
    \label{fig:correctness}
\end{figure*}

\subsection{Effect on the Quality of the Final SQL Query}
Figure \ref{fig:correctness} right-facet shows the average correctness of the final SQL queries by condition. The Kruskal-Wallis H test for the correctness between conditions was not significant ($\chi^2(2) = 3.85$, $df = 2$, $p = 0.14$). However, correctness was directionally higher for both LLM conditions than for the web search condition. 
% > mean_correctness
% # A tibble: 3 × 3
%   Task                 mean_correctness_scaled se_scaled
%   <fct>                                  <dbl>     <dbl>
% 1 Web Search                              4.72      2.16
% 2 ChatGPT                                19.7       8.28
% 3 Instructor-tuned LLM                   19.3       4.99

\subsection{Self-Reported Mental Demand}
The Kruskal-Wallis H test for mental demands was not significant ($\chi^2(2) = 2.05$, $df = 2$, $p = 0.36$). However, there is suggestive evidence that students in the Instructor-tuned LLM conditions reported that the task was less mentally demanding compared to the other conditions (see Figure \ref{fig:correctness} left-facet). When comparing the instructor-tuned LLM with ChatGPT, one of the students commented \textit{``[instructor-tuned LLM] seem more accurate or specific than chatGPT.''} This may indicate lesser mental demand on the student.

% > mean_demands
% # A tibble: 3 × 3
%   Task                 mean_demands    se
%   <fct>                       <dbl> <dbl>
% 1 ChatGPT                      5.17 0.331
% 2 Web Search                   5.24 0.352
% 3 Instructor-tuned LLM         4.5  0.330
\section{Discussion}

\paragraph{Key Findings}
We found that students needed to interact more with the Instructor-tuned LLM compared to ChatGPT and Web Search. This increased interaction could be attributed to the system prompt for the Instructor-tuned LLM, which included sentences such as \textit{``The instructor does not provide the exact answer to the given problem...''}, as well as other guardrails to prevent cheating with LLMs \cite{chen2023gptutor, jurenka2024towards}. One might hypothesize that this would lead to lower grades for students using the Instructor-tuned LLM versus ChatGPT and Web Search, which can readily provide direct answers. However, our results did not show significant differences in the quality of the final SQL queries between conditions. This is promising, as greater engagement could potentially lead to longer-term learning \cite{bastani2024generative, kumar2023impact}. Students expressed interest in using instructor-tuned LLM over ChatGPT and said \textit{``Would rather use [instructor-tuned LLM] over ChatGPT given the prior knowledge of the tables and helps with practical examples of how to join two tables.''} This supports the idea that scaffolded learning, in which students are guided but not given direct answers, can be as effective as direct instruction \cite{hmelo2007scaffolding, maybin1992scaffolding}.

Interestingly, the higher number of interactions with the Instructor-tuned LLM did not result in higher reported mental demand. In fact, students who used Instructor-tuned LLM reported levels of mental demand that were equal to or lower than those using ChatGPT and Web Search. Lowering the cognitive load can make programming more approachable, potentially reducing dropout rates in CS and encouraging more students to pursue and persist in the field \cite{scragg1998study, wang2017diversity}. Additionally, there were no differences in the number of changes made to the final SQL query between the different conditions. This suggests that while the Instructor-tuned LLM may require more interaction, it does not necessarily increase the mental burden on students and maintains the same level of code refinement as other methods \cite{edwards2022practical, bueno2021effects}. In summary, our findings highlight the potential of using LLMs as facilitators of learning rather than just sources of information and contribute to the growing literature of designing pedagogically informed LLM-tutors \cite{jurenka2024towards, kumar2024supporting}.

\paragraph{Broader Implications}

Studies such as this show the value of low-cost instructor tuning through system prompting for increasing student engagement. Instructor-tuned chatbots can serve as “levers” for instructors wishing to amplify student engagement in their courses outside of the classroom by offering personalized support to students through chatbots that have been “tuned” with course-specific context and content. The provision of these chatbots may also reduce over-dependency on general-purpose chat agents like ChatGPT \cite{cotton2024chatting, adeshola2023opportunities, kumar2023math}. At the same time, the effectiveness of these levers will depend on both the accuracy of the models themselves (something we can expect to improve with future frontier models) and the ability of students to ask the right questions (through prompts, which is harder than it may appear to users \cite{10.1145/3544548.3581388, kumar2023impact}). Without the latter, increased engagement may not translate into increased learning, as we saw with the ratings of the quality of the final query in our study. Therefore, providing better support for students by asking the \textit{right} questions is a useful direction for future work. The lower mental demand reported by participants for the instructor-tuned condition also hints at the potential value of instructor-tuning for helping users manage the metacognitive demands of using AI, as found in recent studies \cite{10.1145/3613904.3642902, wu2023integrating}.

\paragraph{Limitations \& Future Work}

% - small sample size; ideally, would have done a between-subject randomized controlled experiment, but due to power considerations, we had to think of a within-subjects design (although everything was randomized, such as the question order, condition order, etc.)
The small sample size presents a primary threat to validity. Although measures such as randomizing questions and condition orders were implemented to mitigate biases, the limited sample size may still impact the generalizability and power of the findings. The participants in our study were upper-year CS students from a research-intensive university. This further raises questions about the validity of the general population's findings related to SQL query writing.
% - measure long-term `learning'
Additionally, the study did not measure long-term learning outcomes, which limits understanding of how the different methods influence sustained learning and retention over time. Moreover, the presence of the interviewer during the programming task may have affected the students' help-seeking behavior \cite{adair1984hawthorne}.

Future work may involve larger-scale, between-subjects, randomized controlled experiments to enhance the generalizability and robustness of the findings, such as through a multi-institutional longitudinal study. Furthermore, investigating the long-term learning outcomes associated with using LLMs versus web search for coding assistance will provide deeper insights into their impacts on sustained learning and retention. Beyond prompting, instructor-tuned LLM can be made even more useful for learning by fine-tuning the language models with pedagogically rich data \cite{jurenka2024towards}. Exploring different types of programming tasks and expanding the study to diverse educational settings could further elucidate LLMs' broader applicability and effectiveness in data systems education. Future work should investigate how much \textit{learning} happens when using LLMs for programming, compared to relying on search engines. One might hypothesize that getting direct solutions from LLMs may hamper learning compared to getting clues from search. In this case, instructor-provided LLMs can hold key in balancing the tradeoffs between the benefits of using LLMs with the amount of learning on the students' part.

\section{Conclusion}
We conducted a randomized interview study to compare students' use of LLMs (out-of-the-box and instructor-tuned) with conventional web search (status quo) for writing SQL queries. Our findings suggest that an instructor-tuned LLM might lead to higher engagement, on average, compared to other sources of help while maintaining the quality of downstream performance on the given task. Preliminary evidence also points to a reduction in students' self-reported mental demand for writing SQL queries while utilizing the instructor-tuned LLM. These findings highlight the potential of designing LLM-based instructional resources with the participation of teachers and have implications for the field of productivity and the future of work, in addition to education technology.
\begin{acks}

% This project was financially supported by REDACTED GRANT \#123456 and REDACTED grant \#123456. We would also like to thank REDACTED PERSON and REDACTED PERSON for their support and guidance.
This work was supported by the Natural Sciences and Engineering Research Council of Canada (NSERC) grant \#RGPIN-2024-04348, and the Learning \& Education Advancement Fund from the Office of the Vice-Provost, Innovations in Undergraduate Education, University of Toronto.

\end{acks}

\balance
\bibliographystyle{ACM-Reference-Format}
\bibliography{main}

% \appendix

% \section{Appendix}

\end{document}